\input amstex
\documentstyle{amsppt}
\magnification=1100
\headline={\hfill\number\pageno}

\topmatter

\hfill {\eightpoint{IASSNS-HEP-97/37}}  

\bigpagebreak

\title \quad \\
A Rejoinder on Quaternionic \\
Projective Representations
\endtitle
\author S.L. Adler ${}^{\star}$ 
and G.G. Emch ${}^{\dagger}$
\endauthor
\affil ${}^{\star}$ Institute for Advanced Study, Princeton, 
NJ 08540 \\
${}^{\dagger}$ Dept. of Mathematics, Univ. of Florida, Gainesville, FL 32611
\endaffil


\nologo

\abstract{In a series of papers published in this Journal, a discussion 
was started on the significance of a new definition of projective 
representations in quaternionic Hilbert spaces. The present paper gives 
what we believe is a resolution of the semantic differences that had 
apparently tended to obscure the isssues. }
\endabstract

\endtopmatter

\document

We must first harmonize the notations in papers 
$ {}^{1,2,3,4} $ 
that were written more than thirty years apart, and for different audiences. 
Let 
$ {\Cal H}_{\Bbb H} $ 
be a quaternionic Hilbert space. In order to facilitate the transcription 
to Dirac's bra--ket notation, we write the multiplication by scalars on 
the right, with the scalar product defined to be linear in its second term: 
$$ ( \psi p , \phi q) = p^* (\psi , \phi) q 
\tag 1 $$
in conformity with
$\vert \phi q > =\vert \phi> q \, . $

Under the initial assumptions of Wigner
$ {}^{5} \, , $
reformulated by Bargmann
$ {}^{6} \, , $ 
or the assumptions of Emch and Piron
$ {}^{7} \, , $ 
a symmetry 
$ \mu $ 
is defined as a map that preserves transition probabilities between rays, or 
equivalently as an automorphism of the ortho--complemented lattice 
$ {\Cal P}({\Cal H}_{\Bbb H}) $
the elements of which are the closed subspaces (i.e. the projectors) of 
the Hilbert space 
$ {\Cal H}_{\Bbb H} \, . $ 

The theorem known as Wigner's theorem (by physicists), and as the infinite
dimensional version of the fundamental theorem 
$ {}^{8} $ 
of projective geometry (by mathematicians) asserts that every symmetry is 
implemented by a co-unitary operator  
$ U $ 
satisfying :
$$ P \in  {\Cal P}({\Cal H}_{\Bbb H}) \quad \Rightarrow \quad 
 \mu [P] = U^* P U  
\tag 2 $$
with
$$ U (\psi q) = (U \psi) \alpha_U [q]  \quad \forall \quad 
\psi \in {\Cal H}_{\Bbb H} \quad {\text and\/} \quad q \in {\Bbb H} 
\tag 3a  $$
and 
$$ 
\alpha_U [q] = \omega_U^* q \omega_U 
\quad \text{ for some} \quad
 \omega_U  \in {\Bbb H}
\quad \text{ with} \quad  
\omega_U^*  \omega_U = 1 \, ; 
\tag 3b $$
i.e.
$\alpha_U $ 
is an automorphism of the field of quaternions. The co-unitarity of 
$ U $ 
means that
$$ U^*U = U U^* = I 
\quad \text{ so that} \quad 
(U \psi , U \phi ) = \alpha_U [(\psi , \phi )] 
\tag 3c $$
which reflects the fact that for a co-linear operator $ A $ the adjoint
is defined by
$$ (A^* \psi , \phi ) = \alpha_A^{-1} [(\psi , A \phi )] \, . 
\tag 4 $$
Conversely, every co-unitary operator implements a symmetry.

Finally, a symmetry determines  the co-unitary operator that implements it,
uniquely up to a ``phase''; specifically the quaternionic form of Schur's 
lemma 
$ {}^{1} $ 
implies that two co-unitary operators 
$ U_1 $ and $ U_2 $ implement the same symmetry if and only if there exists 
a unit quaternion 
$ \omega \, , $ 
such that
$ U_2 =  U_1 C_{\omega} $ 
where
$ C_{\omega} $ is the co-unitary operator defined by
$$ C_{\omega} \psi = \psi \, \omega \quad . 
\tag 5 $$
Indeed, 
$$ P \in  {\Cal P}({\Cal H}_{\Bbb H}) \quad \Rightarrow \quad 
C_{\omega}^*\,  P\,  C_{\omega} = P  \quad .
\tag 6 $$  

Hence for every symmetry separately, one can choose a {\it unitary}
operator to implement this symmetry; and this unitary operator is unique 
up to a sign.

So far, and as long as each symmetry is treated separately, 
the above approach covers the premises of both Adler 
$ {}^{2} $ 
and Emch 
$ {}^{1} \, . $

When an abstract group $ G $ is represented as a group of symmetries, i.e.
when a symmetry 
$ \mu(g) $ 
is assigned to every
$ g \in G $ 
in such a manner that
$$ \mu(g_1) \mu(g_2) = \mu(g_1 g_2) 
\quad \forall \quad
(g_1,g_2) \in G \times G \, ,
\tag 7a $$
i.e.
$$    P \in  {\Cal P}({\Cal H}_{\Bbb H}) \quad \Rightarrow \quad
\mu(g_1)[\, \mu(g_2)[P]] = \mu(g_1 g_2) [P] 
\quad \forall \quad
(g_1,g_2) \in G \times G \, ,
\tag 7b $$
one can repeat the above procedure for each 
$ g $ 
separately, and obtain a lifting by unitary operators 
$ U(g) $ 
satisfying
$$ U(g_1) U(g_2) = \pm \, U(g_1 g_2) \, . 
\tag 8 $$

When 
$ G $ 
is a Lie group, and 
$ \mu $ 
is a {\it continuous} representation, the brutal lifting just described may, 
however, not lead to a {\it continuous} unitary representation. As physics 
needs continuity to define the observables corresponding to the generators of 
the unitary representation, it is reassuring to know that continuity obtains
nevertheless 
$ {}^{1} \, , $ 
as a result of the following procedure.

Firstly, one shows that there always exists a continuous local lifting 
by co-unitary operators, satisfying thus the condition

$$ U(g_1) U(g_2) = U(g_1 g_2)\, C_{\omega(g_1,g_2)} \quad .
\tag 9 $$
In this expression 
$ C_{\omega} $ 
is a co-unitary operator, defined as in (5), where now 
$ \omega = \omega(.,.) $ 
is a continuous function of each of its arguments, takes its values in the 
unit quaternions, and satisfies, besides the trivial conditions
$ \omega(g,e) = \omega(e,g) \, , $ 
the 2-cocycle condition:
$$ \omega(g_1 , g_2 g_3)\,\,\omega(g_2 , g_3) \, = \,
\omega (g_1 g_2 , g_3)\,\, \alpha_{U_{g_3}^{-1}}[\,\omega(g_1 , g_2)\, ]\quad ;
\tag 10a $$
for the purpose of ulterior comparison with (13), we rewrite (10a) as:
$$ C_{\omega(g_1 , g_2 g_3)}\,\,C_{\omega(g_2 , g_3)} \, = \,
C_{\omega (g_1 g_2 , g_3)}\,\, {U_{g_3}^{-1}}\, 
C_{\omega(g_1 , g_2)}\, U_{g_3} \quad .
\tag 10b $$

Secondly, one shows that such a lifting is always equivalent to a 
{\it continuous, unitary, local,\/} but true {\it representation\/} (i.e. no 
$ \omega \, , $ 
not even a 
$ \pm $ 
sign, ambiguity). 

Thirdly, whenever the Lie group $ G $  is simply connected, 
this can be extended to a continuous, unitary representation of the whole
group $ G \, . $ In cases where the group is doubly connected (e.g. the
rotation group in three dimensions), one only obtains the above result for 
its covering group; it is when one has to consider the group itself, that the 
$ \pm $ ambiguity of (8) can possibly manifest itself. 
As the latter amendment (covering multiply connected groups) is not germane 
to the issue on which we want to concentrate in this paper, we will not
pursue that part of the discussion here. 

The straightforward generalization we just sketched, extending to  
quaternionic Hilbert spaces the analysis familiar from the complex Hilbert
spaces situation, presents one remarkable feature: the ``phase reduction'' 
is always locally trivial. Mathematically, this can be 
understood 
$ {}^{1} $ 
from the fact that the local phase reduction amounts to finding, 
up to equivalence,  all the extensions 
$ {}^{9} $ 
of the Lie algebra of 
$ G $ 
by the Lie algebra of the group of automorphisms of the field of quaternions; 
as the latter happens to be the semi-simple Lie algebra 
$ su(2,C) \, , $ 
all such extensions are trivial
$ {}^{10} \, . $
In this respect the complex case is much more involved, as shown by
Bargmann
$ {}^{11} \, . $ 
In particular, the phase reduction is {\it not} locally trivial
for the Galilei group, a fact that is interpreted as viewing the mass as 
parametrizing the sectors of a superselection rule. Two attitudes are
possible in this juncture. The first, which was chosen by Emch 
$ {}^{1} \, , $ 
was to accept that Galilean QM {\it is} different in its quaternionic 
realization from what it is in its complex realization. The second is to 
pursue the issue, and to generalize the definition of a projective 
representation; this was recently proposed by Adler 
$ {}^{2} \, . $

Tranlated in the notation of this paper, Adler's proposal 
$ {}^{2} $ 
is to replace condition (9) by the weakened condition
$$ U(g_1) U(g_2) = U(g_1 g_2)\, L_{\Omega(g_1,g_2)} 
\tag 11 $$
where
$ L_{\Omega(g_1,g_2)} $
is the linear operator
$$ L_{\Omega(g_1,g_2)} \psi = 
\sum_k \phi_k\, \omega_k\,(g_1,g_2)\, (\phi_k , \psi) 
\quad {\text with} \quad
\omega_k\,(g_1,g_2)^* \,\, \omega_k\,(g_1,g_2) = 1
\tag 12 $$
and 
$ \Phi = \{ \phi_k  \, \mid \, k= 1, 2, ... \}  $
is a complete orthonormal basis in 
$ {\Cal H}_{\Bbb H} \, , $
the same for all pairs 
$ (g_1,g_2) $
of elements of $ G \, . $
Note that
$$ L_{\Omega(g_1 , g_2 g_3)}\,\,L_{\Omega(g_2 , g_3)} \, = \,
L_{\Omega (g_1 g_2 , g_3)}\,\, {U_{g_3}^{-1}}\, 
L_{\Omega(g_1 , g_2)}\, U_{g_3} \quad .
\tag 13 $$

While \{11,13\} looks somewhat similar to \{9,10b\}, there are major 
differences between these two formulations; our purpose in this paper is to 
delineate sharply the scope and reach of these variations.

Firstly, (9) is a direct consequence of the condition (7). Hence one should
expect condition (7) to be violated by (11). This is indeed the case: see 
(16) below. Recall that (7) is the defining condition for the usual definition
of a projective representation, as 
$ {\Cal P}({\Cal H}_{\Bbb H}) $ 
is the projective space associated to the vector space 
${\Cal H}_{\Bbb H} \, . $ It is in fact equivalent to (9), and it is the 
condition Adler 
$ {}^{2} $ 
refers to as the defining property of a {\it strong} 
projective representation, in opposition to (11), which is equivalent to (16),
and which he introduces as the definition of a {\it weak} projective 
representation. 

Secondly, (9) is a relation among essentially co-unitary operators. 
It is true, 
as we just mentioned, that a powerful theorem 
${}^{1} $
allows to reduce the phases 
and thus to obtain a locally trivial continuous unitary representation, so that
(9) becomes ultimately a relation between linear operators. Nevertheless,
this reduction is not instructive in the present juncture since it is (9) 
itself [not (8)] that serves as a motivation for the extension (11). 
By contrast, (11) is in its very essence a relation between unitary operators; 
in particular $ L $ is a linear
operator (in fact a unitary operator) that involves the choice of a 
complete orthonormal basis 
$ \Phi = \{ \phi_k  \, \mid \, k= 1, 2, ... \} ; $
i.e the focusing on one complete set of commuting observables, or more 
precisely, on a discrete, maximal abelian, real subalgebra 
$$ {\Cal A}_{\Phi} \, = \, 
\{\,\, A : \psi \in {\Cal H}_{\Bbb H} \mapsto 
A\psi = \sum_k \phi_k\, a_k (\phi_k , \psi) \in {\Cal H}_{\Bbb H} \,\,\}
\tag 14 $$
the minimal projectors of which are the projectors 
$ P_{\phi_k} $
on the one-dimensional rays corresponding to each element 
$ \phi_k $ 
of the chosen basis $ \Phi \, . $ 
We denote by 
$ {\Cal P}({\Cal A}_{\Phi}) $
the Boolean sublattice of
$ {\Cal P}({\Cal H}_{\Bbb H}) $
generated by these projectors.

Thirdly, as a consequence of the above remark, whereas the co-linear operators
$ C_{\omega(g_1,g_2)} $ 
in (9) implement the trivial symmetry (see 6) -- and are in particular 
independent of any choice of a Hilbert space basis --
that is not the case for the symmetry implemented by the linear operators 
$ L_{\Omega (g_1,g_2)} \, . $ 
Indeed, we have generically only:
 
$$ P \in {\Cal P}({\Cal A}_{\Phi}) \quad \Rightarrow \quad 
L_{\Omega(g_1,g_2)}^*\, P\, L_{\Omega(g_1,g_2)} = P  \quad .
\tag 15 $$
Hence, the symmetry implemented by
$ U(g_1 g_2) $ 
coincides with the symmetry implemented by
$ U(g_1) U(g_2) $
only for the elements of the distinguished maximal abelian algebra
$ {\Cal A}_{\Phi} \, $
chosen to define the linear operators 
$ L_{\Omega(g_1,g_2)} \, : $
$$  P \in {\Cal P}({\Cal A}_{\Phi}) \quad \Rightarrow \quad
\mu(g_1)[\, \mu(g_2) [P]] = \mu(g_1 g_2) [P] \quad \forall \quad
(g_1,g_2) \in G \times G \quad .
\tag 16 $$ 
This, compared to (7), is the major difference between the conditions defining
weak vs strong projective representations. While both require, for each 
symmetry {\it separately}, that 
$ \mu (g) $ 
be an automorphism of the {\it whole} system (a condition necessary to 
support the use of Wigner's theorem), the difference appears when it comes 
to the representation of a {\it group} of symmetries: the strong definition 
requires (7b), i.e. that 
$ \mu $ 
is a representation on the full
$ {\Cal P}({\Cal H}_{\Bbb H}) $
whereas the weak definition requires only (16), i.e. that this condition hold 
on
$ {\Cal P}({\Cal A}_{\Phi}) \, . $ 

This is the price one must be prepared to pay for the relaxing from the 
``strong'' condition (9) to the ``weak'' condition (11) 
-- which is the generalization proposed by Adler
$ {}^{2} \, . $
At this price, it has become possible 
$ {}^{12,4,13} \, : $
to classify the irreducible weakly projective representations of connected 
Lie groups; to embed complex projective representations into weakly 
projective quaternionic representations (even when the Bargmann complex 
phase reduction is not locally trivial); to construct quaternionic coherent 
states (including the weakly projective case); and to discuss how, in the 
complex case, the weak condition (11) 
already implies the stronger condition of (9). 

After comparing their original
motivations, the authors realized how they both had hoped to take advantage
of the $ SU(2) $ symmetry of the quaternions: Emch 
$ {}^{1} $
was interested in finding some natural
coupling between the inhomogenous Lorentz group of special
relativity and the internal symmetries then known in elementary particle
theory; Adler 
$ {}^{2} $
was similarly interested in finding a source in the ray structure of 
Hilbert space for the 
color symmetry. It seems fair to say that, even with the generalization 
proposed by Adler
$ {}^{2}\, , $ 
the structure of the current quaternionic models for quantum theories is not 
(yet) rich enough to accomodate dreams that extend beyond the complex Hilbert 
space formalism.

\medpagebreak

\flushpar {\bf Acknowledgements}
 
\medpagebreak

\flushpar The authors thank Dr. A. Jadczyk for discussions on  
matters related to this paper.  The work of S.L. Adler was supported in 
part by the Department of Energy under Grant $\#$DE-FG02-90ER40542.

\enddocument

\end